A search and modeling of peculiar narrow transient line components in novae spectra

Takeda, L. and Diaz, M. P.

IAG, Universidade de São Paulo, Rua do Matão, 1226 São Paulo, SP, 05508-900, Brazil

larissa@mail.com, marcos@astro.iag.usp.br

Abstract. In this paper we discuss the formation of peculiar narrow emission line components observed in the spectra of a few novae. We aim to identify the physical source responsible for transient narrow components that present orbital radial velocity modulations, first observed in the post-outburst recombination lines of Nova U Sco 2010. A search for candidate novae showing similar narrow components is presented. Photoionization simulations indicate that the forming region cannot be restricted to the primary Roche Lobe, but could be located around the outer Lagrangian point $L_3$. Further analysis disfavors an origin at the accretion disk and the companion star. In addition, we analyze possible correlations between the presence of the narrow components, the basic nova parameters and the spectral classification in the initial permitted phase.

1. Introduction

The analysis of nova spectral evolution is a powerful tool to study the physical properties of nova systems. Williams (1992) found correlations between the initial phase of nova spectra and their late nebular phase, presenting the physical features associated to them. Novae spectra show highly structured broad emission lines during decay. Those structures reflect a complex combination of optical depth, emissivity and velocity field (e.g. Shore et al. 2011). In the permitted lines phase, a few novae present a peculiar narrow emission component overlapping a broad one in their spectra. This component appears in the first days after maximum and has FWHM of a few hundred km/s. For instance, $\Delta t = 3t_3$ and FWHM ~ 500 km/s for U Sco (Diaz et al. 2010). The curious characteristic of these components is that they present orbital radial velocity modulations (Mason et al. 2012), which implies that the emission region shares kinematic properties with the binary system.

Narrow components in nova outburst spectra are not uncommon and many static shell models have been made to explain them. Some novae, especially the Fe II spectral type ones, present mainly narrow components in their spectra. Those emissions are often associated with slower gas expansion velocities in the nova eruption. Different geometries can also reproduce narrow line profiles, as Munari et al. (2010) have shown in their model for V2672 Oph (Nova Oph 2009). They simulated a prolate structure for the nova system composed by polar blobs and an equatorial ring. The radiation emitted by the ring formed a narrow component in the modeled spectra that fitted the observed one.

Possible narrow lines emitting regions connected to the orbital system are the accretion disk and the companion star. In fact, these two regions were proposed to explain the narrow lines in U Sco spectra. Diaz et al. (2010) speculated that the forming region relies on the secondary star, although lacking support from further simulations. On the other hand, Mason et al. (2012) suggested that it was actually the accretion disk that was responsible for the narrow emission. In this paper, we approach this subject with more detailed data analysis and new photoionization calculations to constrain the possible emitting regions.

Despite the fact that narrow components are frequently observed during nova outbursts, they are not expected to present orbital velocity modulations in the early spectra. At this stage, the receding nova pseudo-photosphere is larger than the primary system, at least for CVs with main-sequence companions. Therefore, the radiation with orbital information should not be visible in the first days after outburst.

We also search for nova spectra in 3 synoptic nova surveys trying to identify more objects that present suspicious narrow line components. Then, we compare the presence of the narrow components in classical and recurrent nova spectra and analyze the differences and similarities found. Possible correlations of basic nova properties with the presence of the narrow components were also surveyed.

## 2. The search for transient narrow components

### 2.1. Synoptic Observations

Using the optical spectral data from the CTIO Nova Survey (Williams et al. 1991, 1994), the SOAR Telescope Synoptic Survey of Novae and the Stony Brook/SMARTS Atlas of (mostly) Southern Novae (Walter et al. 2012), it was possible to identify nova candidates that present transient narrow line components.

The synoptic nova surveys are essential sources of spectral data in the present analysis, because the nova eruptions are observed and followed as long as possible. One of the problems we found in our search was the lack of time coverage in most of spectral observations. As the components we are trying to locate are transient, it was very difficult to find the right spectra at the right time. It is likely that we have missed these components in many objects, what affects directly our statistical analysis (see section 3).

Almost all of the analyzed spectra were obtained at the 3 catalogs mentioned above. The CTIO Nova Survey contains data of nova eruptions from 1987 to 1994. The spectra have broad wavelength coverage, but low spectral resolution (between 5 and 16Å). The SOAR Telescope

Synoptic Survey of Novae was active between 2007 and 2011, and has followed the spectral evolution of 16 novae. The wavelength coverage is slightly smaller than the CTIO Nova Survey, and the spectral resolution is about 2 to 3Å. The Stony Brook/SMARTS Atlas of (mostly) Southern Novae is the most recent project – it started in 2003 and it is still in operation. Its data is public and covers more than 60 nova eruption spectra, with extended time coverage and spectral resolution from 0.1 Å to 17.2 Å.

For better understanding of the narrow profile variations, we need both time resolved and moderate to high-resolution spectral observations of novae that present the narrow emission components. Unfortunately, this is very difficult to obtain, because the most probable candidates are the very fast novae, as shown in section 3. By the time we identify the narrow component, we would not be able to get the necessary time coverage.

2.2. Narrow component selection

The recurrent nova U Sco 2010 is the prototype object (Diaz et al. 2010; Mason et al. 2012) roughly defining the expected width and intensities of these structures. We identified the narrow component velocity modulation in this nova spectra and it was later confirmed by Mason et al. (2012).

In our survey, we searched for the presence of narrow emission line components with FWHM in the interval of 500 to 1000 km/s over a broader profile in their early permitted spectra. The narrow components seem to appear in the recombination lines at epochs ranging from 1 to 5 $t_3$. As the objects presenting the narrow components are mostly fast novae (see section 3), this period is no longer than 30 days. We also tried to obtain radial velocity values when there were high-resolution spectra available. For relative velocity measurements, we used the interstellar medium Na I D absorption lines as wavelength reference.

We also identified the narrow component in semi-forbidden lines in IUE spectra of V394 CrA 1987 (Starrfield et al. 1988), suggesting that the density of the forming region is compatible with the formation of narrow recombination and intercombination lines.

3. Statistical Analysis

We analyzed the spectral evolution of 78 novae (11 recurrent and 67 classical). For 21 novae, we did not have enough time coverage, so it was not possible to rule out or confirm the presence of the narrow components. From the remainder (Table 1), V394 CrA, Nova LMC 1990-2, U Sco, Nova LMC 1988-2, Nova LMC 2009, Nova Oph 2009c, DE Cir, V444 Sct, KT Eri, YY Dor, Nova Sco 2007 and Nova Sco 2011b fit the selection criteria and are considered candidates to present the peculiar transient narrow components. There are 5 confirmed recurrent novae that present the narrow system, in contrast to 7 classical ones. We stress that KT Eri is suspected to be a recurrent nova, but until it is confirmed, we will consider it as a classical one (Jurdana-Šepic et al. 2012). Therefore, the narrow component is more likely to appear in recurrent novae. Examples of the nova H$\alpha$ profiles with narrow components over a broad line base are shown in figure 1.

In this sample, only U Sco and Nova Sco 2007 are confirmed to present radial velocity modulations in the narrow components. Besides the instrumental difficulties, the inclination of the orbital system may also hamper the radial velocity measurement, since they would not be detected at high inclinations. Therefore, if there are other cases among novae, actual detections may represent a small subsample of a more frequent phenomenon.

For some of the novae exhibiting the narrow system, we measured the broad to narrow line flux ratio for the most prominent lines. The results are shown in figure 2. The ratio presents a fast increase for all selected novae. For U Sco, it is possible to see this ratio decreasing as the narrow components completely merge into the broad ones. The narrow component fluxes can reach from $0.1F_{broad}$ to $3.5F_{broad}$, and they do not seem to have a preferential scale to the broad component fluxes. Lower flux ratios may be present, but they are difficult to identify.

We also searched for correlations between the emergence of the narrow system and the following basic nova properties: decay time, orbital period, absolute magnitude at maximum and spectral type according to Williams classification. Figure 3(a) shows the distribution of decay time among the studied novae, emphasizing the ones with narrow components. The majority of the selected novae are classified as fast or very fast novae, with one exception: Nova Sco 2007 (V1280 Sco). The actual decay time of this nova is uncertain due to the massive dust formation in the early phase of the outburst (Naito et al. 2012), but it is considered a slow nova. Therefore it is possible that the presence of the transient narrow emission line component was related to the speed class. But Nova Sco 2007 spectra also presented these components with confirmed radial velocity modulation, even being a slow nova. Figure 4 shows the evolution of H$\delta$ line in V1280 Sco in contrast to the He II line in U Sco (2010). The broad components velocities are higher in the U Sco case, which is similar to the other selected candidates. The narrow component velocity modulations are also higher in the U Sco spectra (Mason et al. 2012).

Figure 3(b) shows the distribution of outburst amplitude of the studied novae. The data for the candidates with narrow components is not complete because there are no observations at quiescence available in the literature for some of the surveyed novae in LMC. It is possible to observe in the histogram that the average amplitude for all novae is higher than for the novae with narrow components. This may be due to the fact that recurrent novae, the category that most candidates belong to, usually have smaller amplitudes than classical novae (Warner 1995). In a comparison between the narrow component novae and the recurrent novae, we obtained 50% of match for the amplitude distribution and 84% for the $t_3$ distribution, using a Kolmogorov-Smirnov test.

Another test suggests that the decay time distributions for all novae and for the sample of novae with narrow line components are not compatible, with a matching probability of 5.1%. For the amplitude case, the distributions also don't match with P = 10%. Even considering the small number statistics (that are only 12 novae with the narrow components in the sample), the comparison indicates that these components are more common among fast novae.

The spectral type analysis revealed that 9 of the novae exhibiting narrow profiles are He/N novae. Nova LMC 1988-2 and V444 Sct (Nova Scuti 1991) are classified as Fe IIb type (Williams 1992), a class related to the He/N type. The only Fe II type nova is V1280 Sco (Nova Sco 2007). Further correlations could not be found with the data at hand.

4. Photoionization modeling

4.1. Closed geometry models

In order to identify the physical source of the narrow emission components, we modeled three novae using the photoionization code RAINY3D (Moraes & Diaz 2011), which runs CLOUDY v06.02b (Ferland et al. 1998) as a subroutine. We evaluated if the observed narrow line fluxes could be reproduced considering only the emission from the gas inside the primary Roche Lobe. Simulations of the novae U Sco in the 2010 outburst, V394 CrA in 1987, and KT Eri in 2009 were performed. For all simulations, we assumed the lines were produced in a spherical cloud with homogeneous gas and covering and filling factor 1. These choices were intended to maximize the recombination line emission from photoionized gas inside the RL. We estimated the Roche Lobe radii using the orbital periods and the Eggleton relation (Eggleton 1983). The closed geometry was chosen to simulate the case of the ionizing source inside the emission region, allowing that the cloud's radiation towards the central source would interact with the gas at the opposite side. The models were calculated using Rauch Stellar Atmosphere catalog and mean chemical composition for novae, when we could not find specific information about it. The input parameters are displayed in Table 2.

For the V394 CrA models, we used the spectral data from CTIO Nova Survey. The spectrum at day 28 after maximum was deredden with $E(B-V) = 1.10$ (Hachisu & Kato 2000). The luminosity of the central source was limited to the range $36 < \log(L) < 38$ erg/s, as there were no X-rays data available. The model luminosities were scaled to a distance of 4.2 kpc (Hachisu & Kato 2000).

In the case of KT Eri, the models were performed with Stony Brook/SMARTS Atlas spectral data. The line fluxes were measured at day 22 after maximum, after the reddening correction with E(B−V) = 0.08 (Ragan et al. 2009). As for V394 CrA, the central source luminosity is constraint to the range 36 < log(L) < 38 erg/s. The distance adopted was 7 kpc (Imamura & Tanabe 2012). The orbital period of this system is controversial with an indication of a short period of 56 days determined by Hung et al. (2011)

The U Sco models were based on the Δt = 9 d spectra from SOAR Telescope Synoptic Survey of Novae. The reddening was corrected with E(B−V) = 0.15 (Diaz et al. 2010). The central source luminosity was limited to the values obtained from the X-rays observations (Schlegel et al. 2010). The distance was 7.5 kpc (Hachisu et al. 2000).

Simulations with clumpy gas were also made for U Sco, aiming to evaluate the impact of inhomogeneous gas emission. The gas globules were set to 30%, 70% and 80% of the total shell mass, with 10% to 80% of the shell radius, distributed in a random Gaussian function (Diaz et al. 2010).

The results are displayed in figures 5 and 6, as a comparison of the observed line fluxes and the model ones. The symbols in the top graph represents the Hα line fluxes as a function of the hydrogen density in the Roche Lobe, and the symbols in the bottom one, the sum of all narrow line fluxes measured also as a function of the hydrogen density. The thick black lines correspond to the measured line fluxes in the spectra, assuming high measurement errors of ~ 40%. These errors arise mostly from the deblending of the narrow component in the line profile.

All simulations suggest that the narrow line fluxes from the models do not reach the observed values. As the hydrogen densities increase, the gas becomes optically thick before the line flux emissivity could achieve the observable ones. Furthermore, the errors in distance, E(B−V) , luminosity of the central source and the star's mass ratio do not explain the mismatch. We tested a plausible range of values for these parameters, and the results were very similar to those presented in figure 5. Still centering the emitting cloud in the primary star, we also evaluated the necessary volume of gas for the modeled line fluxes match the observed ones. The result was that the radius of the emitting region should be at least 10 times the Roche Lobe for a match. If a smaller binary system is considered for KT Eri ($P_{orb}$ = 56 days), the model luminosities are decreased.

4.2. Open geometry models

Since the grid of closed geometry models fail to reproduce the required luminosities, we simulated a density enhanced cloud in the form of a spherical thick shell sector. The sector comprises a solid angle of $\pi$ (sr) as seen from the central source centered in the outer Lagrangian point $L_3$ (Kopal 1959, 1978), and illuminated from outside by the nova photosphere. The volume of the cloud was restricted so it wouldn't enter the primary Roche Lobe. The other outer Lagrangian points ($L_2$, $L_4$ and $L_5$) were not considered in this work due to the fact they are either possibly too unstable to hold a density enhancement, or hidden by the secondary star (see further discussion in the next section).

We performed the models using CLOUDY's open geometry case, assuming typical values of mass, temperature and luminosity for the central source. The volume and hydrogen densities of the cloud were varied as shown in figures 7 and 8. The H$\alpha$ luminosity results are displayed in the figures as a function of the hydrogen density and volume. The shaded area indicates the H$\alpha$ luminosity measured in the U Sco (2010) spectra, using the distance of 7500 pc. We assumed high measurement error bars (of 40%) again.

Unlike the closed models of the Roche Lobe gas, the emission from a cloud centered in the outer Lagrangian point are able to match the observed narrow line component fluxes. The volume is large enough for the luminosity produced be consistent with the observed data. The hydrogen density of these modeled clouds are still high, in the interval $10^{10} < n_H < 10^{11}$ cm$^{-3}$, but the environment is optically thin, so the lines are not completely absorbed by the cloud. Clouds with lower densities as $n_H = 10^9$ cm$^{-3}$ also could reproduce the observed radiation fluxes, but they would be centered slightly further away from $L_3$.

5. Discussion

We have selected 12 novae exhibiting narrow line components with similar features. Only in U Sco and Nova Sco 2007 spectra it was possible to measure radial velocity modulations, but we suspect that at least for the U Sco type RNe, the line forming process is the same, given the similar FWHM and the time after outburst of their appearance. It is also possible that the manifestation of these narrow components is related to the speed class - fast and very fast novae - and spectral type - He/N or Fe IIb. In this case, Nova Sco 2007 would represent an exception, which could be interpreted as a different phenomenon. We also searched for correlations between the presence of the narrow components and the shape and irregularities in the eruption light curve, but we could not find any relevant feature in the light curve during the narrow component phase.

To find new novae presenting the narrow feature, it is necessary to make high-resolution spectrophotometric observations with time coverage from $t_3$ to $\sim 5t_3$. Time-resolved spectra of the same object over a period of hours would make it possible to measure the orbital velocity of the narrow components with precision and eventually map its origin using Doppler tomography techniques. The narrow component profile and its radial velocity modulations (namely its phase and amplitude) depend on intrinsic (outflow) and orbital motion of the photoionized gas. It is not possible to isolate such components with the data available in the literature.

The observations of the narrow lines in IUE spectra (namely N IV] 1488Å, C IV 1548Å, He II 1640Å and O III] 1663Å) suggest that the density of the region is proper for intercombination lines formation. The usual densities of the accretion disk are too high to allow intercombination line formation. In addition, the component FWHM and FWZI are possibly too small to be formed in an accretion disk.

Our models depict a circunbinary matter bounded photoionized gas with high effective recombination coefficients. The recombination line emissivity scales roughly with the square of gas density and model densities are close to the optically thin maximum values. The model region volume is much larger than a disk atmosphere or the companion irradiated chromosphere, even if an optically thin disk is considered. Therefore, the model emission measure is larger than a disk or chromosphere, where the Balmer lines are formed by recombination plus collisionally excited transitions, with the latter depending linearly on the neutral hydrogen density. The hydrogen line luminosities of bright accretion disks in CVs are well constrained by observations. For instance, the high M-dot disk in V3885 Sgr (a nova-like system with HIPPARCOS parallax distance of 110 pc) has $L_{H\alpha} \sim 7 \times 10^{29}$ erg/s (Ribeiro & Diaz 2007). This is roughly 3 orders of magnitude lower than the estimated narrow transient component luminosities.

The photoionization simulations showed that the forming region should not be restricted to the Roche Lobe, as its volume is not enough to reproduce the observed narrow line component fluxes. These model fluxes reinforce the fact that the disk and the secondary wouldn't be able to reproduce the observed fluxes either. Despite the simplicity of these models, any added structure, as clumps or condensations for instance, would decrease the gas emission which, in most cases, is already lower than the observed data.

Results from open geometry models were consistent with the observed fluxes of U Sco (2010), indicating that the forming region could be such a cloud around the outer Lagrangian point $L_3$. We are aware that these models are too simple to describe the real geometry and kinematics of the gas around $L_3$. Given the radial velocities observed and the results that exclude the structures with volume restricted by the RL, the outer Lagrangian points are the simplest candidates. However, there is no direct information about the outflow density distribution and contrast around the Lagrangian points. Detailed simulations of quiescent mass loss by Bisikalo (2010) suggest that the potential around the semi-stable $L_3$ point is capable of enhance the gas outflow density around $L_3$ to values even higher than our assumptions. The exact shape, size and kinematics of the density enhanced region around $L_3$ during the narrow transient phase are model-dependent. Once the nova pseudo-photosphere recedes inside the RL it may be eventually possible to see both the usual nova remnant continuum emission and the transient narrow components simultaneously. On the other hand, we expect to observe some orbital modulation in X-rays due to absorption by the density enhanced region in a high inclination system like U Sco. Phase-resolved X-ray observations are required to disentangle the absorption from this region from other absorption regions inside the primary RL that have already been proposed by Ness et al. (2012). Most X-ray observations of U Sco are concentrated around the eclipse. Nevertheless, an additional absorption step at phase ~0.6 is mentioned by Ness et al. (2012). This feature may be eventually associated with our proposed emission region ($L_3$), although confirmation from additional cycles would be required to exclude the possibility of unrelated X-ray variability.

There is also a possibility of finding line enhancements at $L_4$ and $L_5$. If these regions are stable enough to increase the density of outflowing gas, there is no reason why the emission would occur only in one side. Emission in both sides ($L_4$ and $L_5$) would produce double peaked components in the spectral lines. High-resolution spectra should provide the data to confirm the existence of these double peaked narrow components with opposite radial velocity phases. Knowing the precise density distribution outside the primary Roche Lobe depends on hydrodynamic simulations of the flow, considering its specific angular momentum and radiation wind acceleration.

The identification of the narrow components in nova spectra requires a minimum spectral resolution. Therefore, it is probable that many nova eruptions have presented these components, but they could not be distinguished from the broad emission lines.

Among the observations that were published in the past decades, there is at least one situation where early orbital profile variations are described. That is the case of the very fast nova V 1500 Cygni (Hutchings & McCall, 1988). V1500 Cyg is a magnetic nova and has profile variations observed at $\Delta t = 3t_3$. This nova has already been compared to U Sco type novae and it presents narrow components that could be result of the same phenomenon responsible for U Sco narrow components. It is not clear if the magnetic field is involved in the narrow line formation in V1500 Cyg. For the novae discussed in this paper, there are no indications of highly magnetic white dwarfs.

6. Conclusions

We found 11 additional novae presenting narrow components similar to U Sco case, including 5 other recurrent novae. It was not possible to measure velocity modulations in their spectra due to the lack of high-resolution data and time coverage, but we believe that for U Sco type novae, the phenomenon that forms these narrow components is possibly the same. Similar novae are good candidates for time resolved high-resolution spectroscopy during early decay.

We used 3 peculiar novae (U Sco, KT Eri and V394 CrA) to investigate this peculiar feature that appears in few other novae. Photoionization simulations have excluded the gas inside the Roche Lobe as the forming region of these lines. We propose and modeled a density-enhanced region at the Lagrangian point $L_3$ as an alternative emitting source. While the density enhancement around this point is a phenomenon that, in theory, may happen in a variety of outflow conditions, the simulations we ran correspond to a simplified geometry that could be eventually applied to other fast novae with similar basic parameters. However, we do not know for sure which and when novae develop such flows and in which ones it would produce observational signatures. We also recognize the limitations of the simplified geometry used in the non-hydrodynamical photoionization models and cannot discard the presence of other sources contributing to this emission.

This research was based on data obtained at the SOAR telescope. We acknowledge the support provided by FAPESP under grant 2012/10533-3 and by CNPq under grant #305725.

Figure captions

Figure 1. Spectra in the Hα region of 5 novae that present the narrow emission line phase, taken from CTIO Nova Survey (Williams et al. 1991, 1994) and SOAR Telescope Synoptic Survey of Novae. The systemic velocity was removed.

Figure 2. Evolution in time of narrow to broad line component flux ratio for selected recombination lines.

Figure 3. Panel (a) shows the decay time distribution for the novae that presented narrow line components in their spectra, in contrast to all surveyed novae (with any spectral data available) and to all novae distribution (with or without spectral data available). Panel (b) shows the amplitude distribution for the novae that presented narrow line components in their spectra, in contrast to all surveyed novae (with any spectral data available) and to all novae distribution (with or without spectral data available).

Figure 4. On the right, high-resolution spectral evolution for Hδ line in V1280 Sco, taken from Stony Brook/SMARTS Atlas (Walter et al. 2012), showing velocity modulations of the narrow component. The dotted line represents the centroid of Hδ emission line in the spectra of July 1$^{st}$ in 2009. The chart on the left shows the He II (4686Å) line in U Sco spectra in 2010, taken from SOAR Telescope Synoptic Survey of Novae. The dotted line represents the centroid of He II on March 19 spectra.

Figure 5. Modeling results for V394 Cra 1987 at Δt = 28 days, U Sco 2010 at Δt = 9 days and KT Eri 2009 at Δt = 22 days. The dot, the plus and the triangle symbols for V394 CrA represent the models for log(L) = 36, log(L) = 37 and log(L) = 38 erg/s, respectively. The dots and the pluses for

U Sco represent log(L) = 36.3 and log(L) = 37.3 erg/s. The dots, the pluses and triangles for KT Eri represent the simulations for log(L) = 36, log(L) = 37 and log(L) = 38 erg/s. The thick black lines indicate the observed fluxes, with error bars of 40%.

Figure 6. Modeling results for U Sco (2010), adopting clumpy gas. The triangle and the plus symbols correspond to gas globules of 30% and 80% of the shell mass, respectively. The black lines indicate the observed fluxes, with error bars of 40%.

Figure 7. Results of the simulations with open geometry indicated by the H$\alpha$ emitted luminosity as a function of the hydrogen density. The shaded region is the measured value of H$\alpha$ narrow component luminosity in U Sco spectra, with assumed errors of 40%. The cloud is centered in $L_3$.

Figure 8. Results of the simulations with open geometry. The H$\alpha$ component luminosity is shown as a function of the product of squared hydrogen density and volume of the cloud. The shaded area is the measured value of H$\alpha$ luminosity in U Sco spectra, with assumed errors of 40%. The cloud is centered in $L_3$.

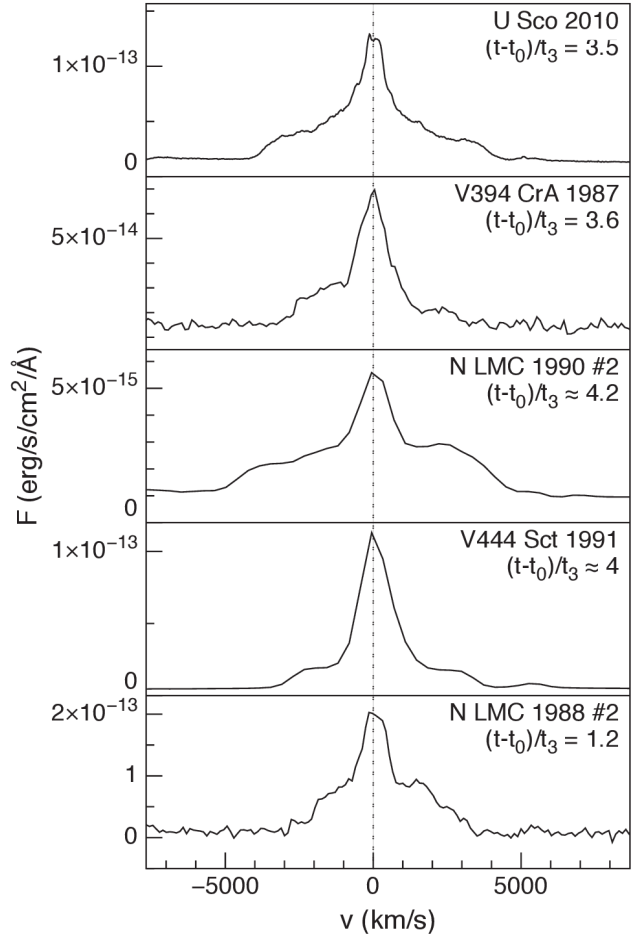

Figure 1

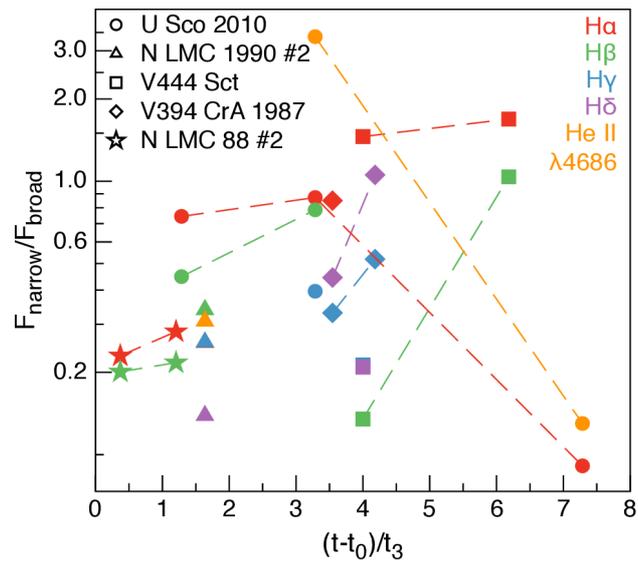

Figure 2

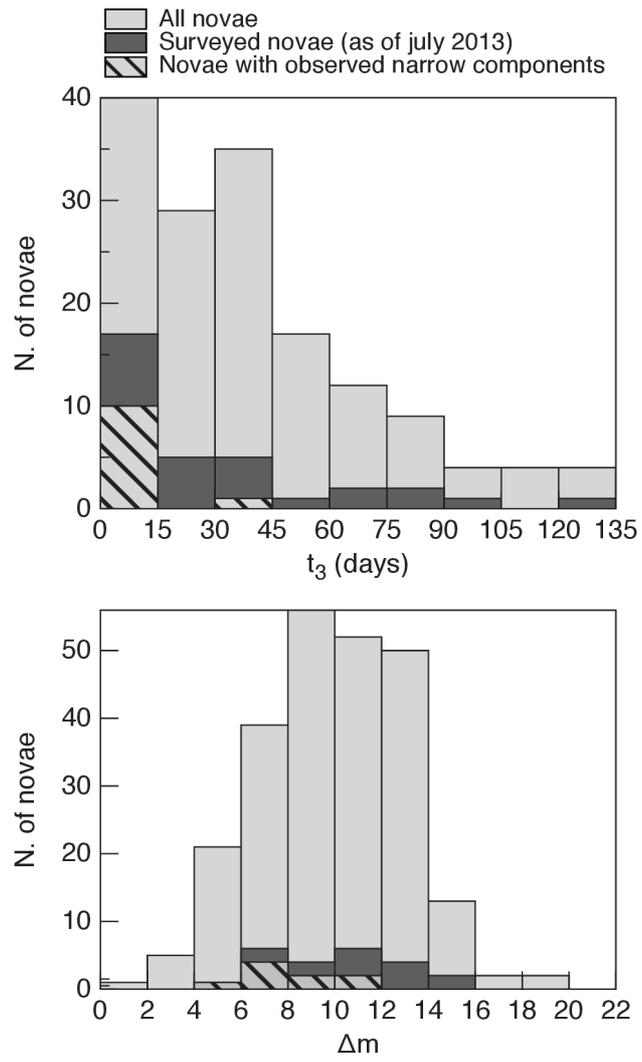

Figure 3

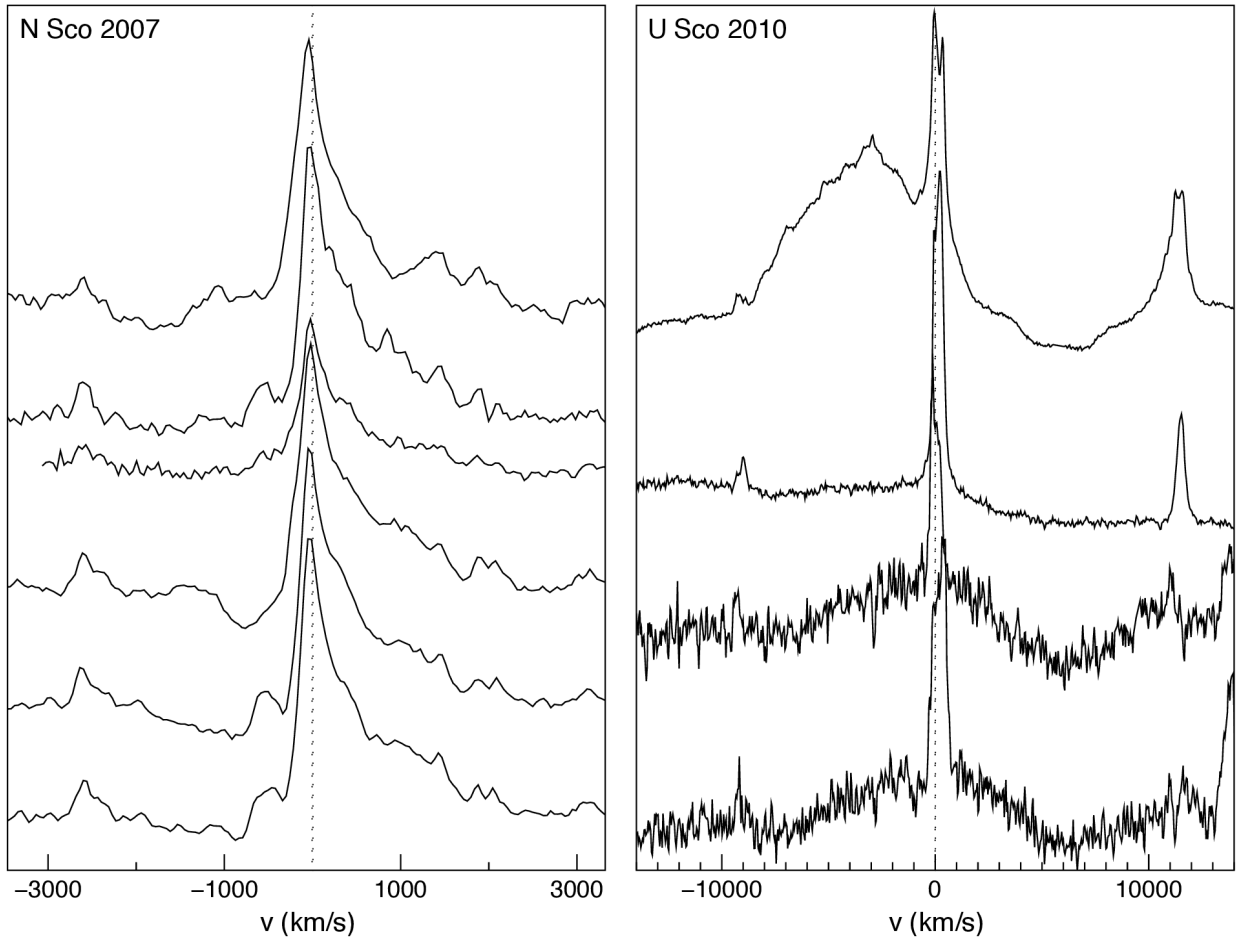

Figure 4

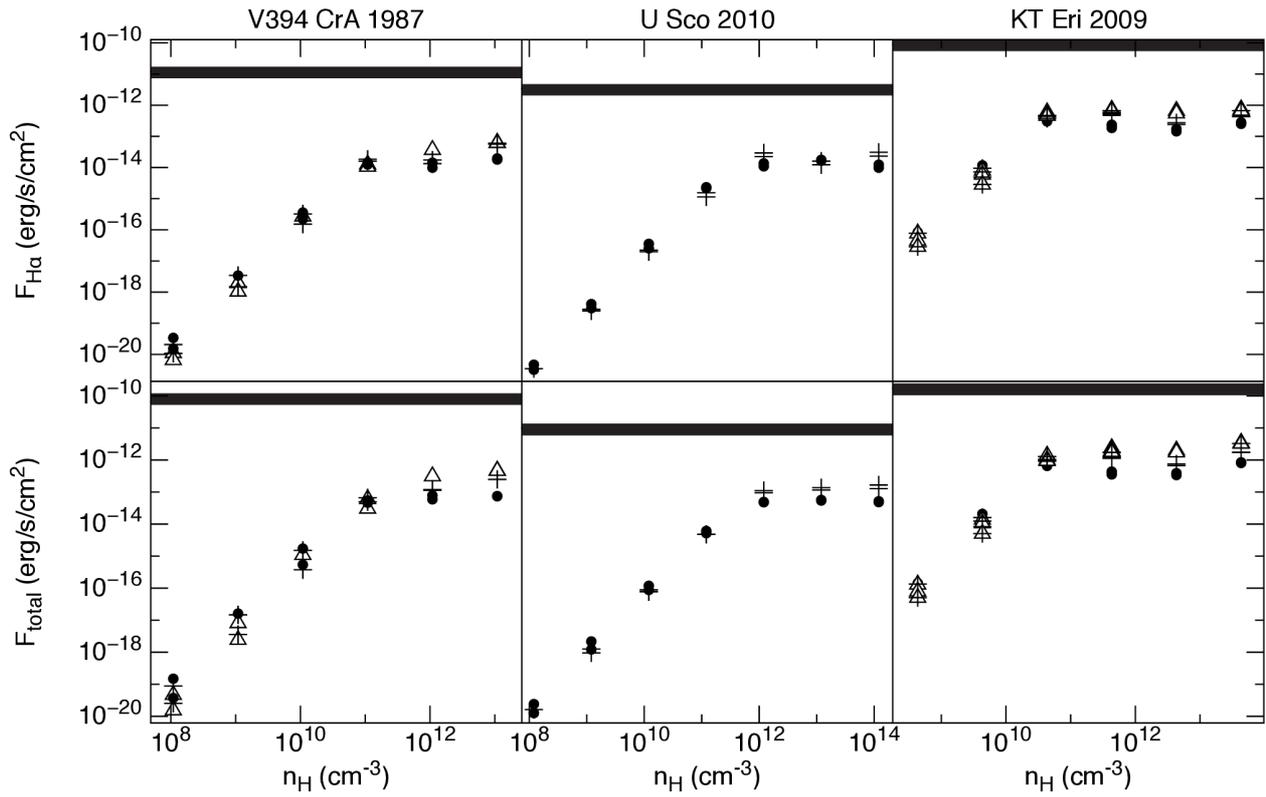

Figure 5

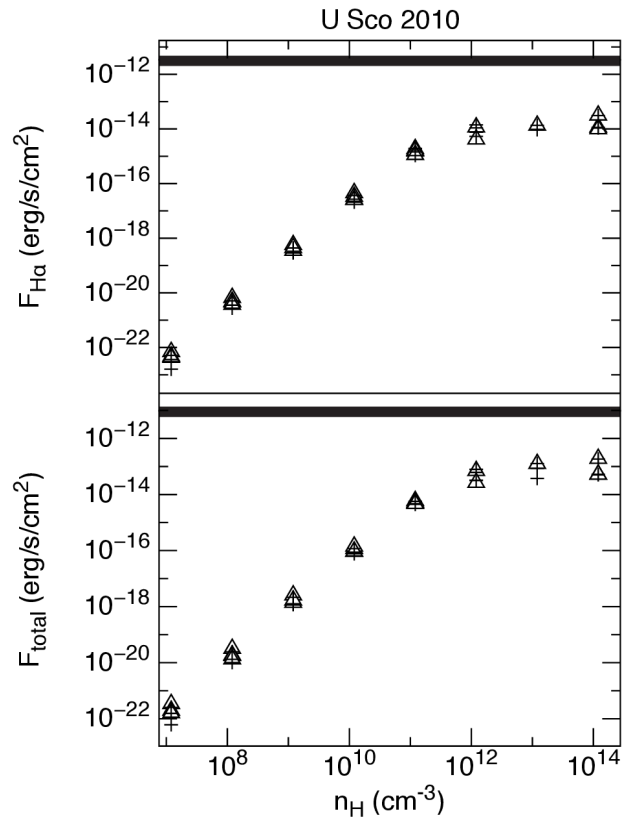

Figure 6

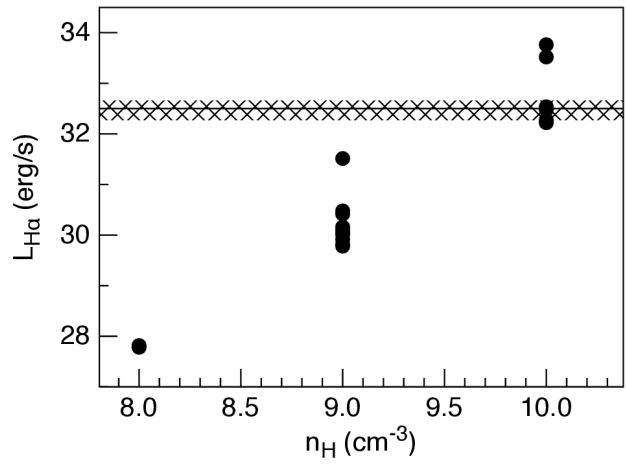

Figure 7

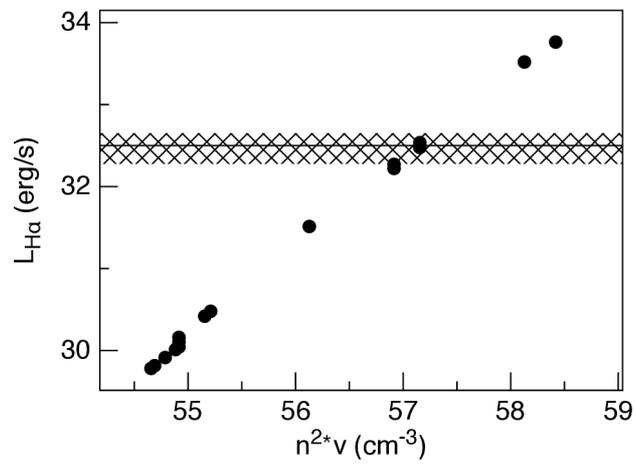

Figure 8

# Table 1

Nova basic parameters, presence of transient narrow components in their spectra and measured orbital modulations (for the studied novae with available spectral data)

| Nova | Type | Observed eclipses | $V_{max}$ | $t_2$ (d) | $t_3$ (d) | $P_{orb}$ (h) | Presence of narrow components | Measured radial modulations |
|---|---|---|---|---|---|---|---|---|
| N Aql 2005 (V1663 Aql) | Classical | No | | | 64 | | No | No |
| N Car 2008 (V679 Car) | Classical | No | | | 127 | | No | No |
| N Car 2009 | Classical | No | | | | | No | No |
| N Cen 2007 (V1065 Cen) | Classical | No | | | 361 | | No | No |
| N Cen 2009 (V1213 Cen) | Classical | No | | | 300 | | No | No |
| N Cen 2012a | Classical | No | | 16 | 34 | | No | No |
| N Cen 2012b | Classical | No | | 12.3 | 19.8 | | No | No |
| V868 Cen | Classical | No | 10.2 | 55 | | | No | No |
| N Cir 2003 (DE Cir) | Classical | No | | | 10 | | Yes | No |
| V394 CrA | Recurrent | Yes | 7.2 | | 5.5 | 36.48 | Yes | No |
| KT Eri | Classical | No | 5.4 | 6.2 | 14.3 | 17688 | Yes | No |
| V838 Her | Classical | No | 5.4 | 2 | 5 | 7.1438 | No | No |
| N LMC 1937 (YY Dor) | Recurrent | No | | 4 | 10.9 | | Yes | No |
| N LMC 1988b | Classical | No | 10.5 | 5 | 10 | | Yes | No |
| N LMC 1990b | Recurrent | No | 10.9 | | 5.5 | 12240 | Yes | No |
| N LMC 1991 | Classical | No | | 5 | 8 | | No | No |
| N LMC 2005 | Classical | No | | | | | No | No |
| N LMC 2009 (N LMC 1971b) | Recurrent | No | 13 | | | 28.56 | Yes | No |
| N LMC 2009b | Classical | No | | | | | No | No |
| N LMC 2012 | Classical | No | | 1.1 | 2.1 | | No | No |
| N Mus 2008 (QY Mus) | Classical | No | | | | | No | No |
| N Nor 2005 (V382 Nor) | Classical | No | | | | | No | No |
| N Nor 2007 (V390 Nor) | Classical | No | | | | | No | No |
| N Oph 1898 (RS Oph) | Recurrent | No | 6.5 | | 10 | 10937 | No | No |
| N Oph 2004 (V2574 Oph) | Classical | No | | | | | No | No |
| N Oph 2006b (V2576 Oph) | Classical | No | | | | | No | No |
| N Oph 2007 (V2615 Oph) | Classical | No | | | 75 | | No | No |
| N Oph 2008 (V2670 Oph) | Classical | No | | | | | No | No |
| N Oph 2008b (V2671 Oph) | Classical | No | | | | | No | No |
| N Oph 2009c (V2672 Oph) | Classical | No | | 2.3 | 4.2 | | Yes | No |

| Name | Type | | | | | | | |
|---|---|---|---|---|---|---|---|---|
| N Oph 2012 | Classical | No | | | | | No | No |
| N Oph 2012b | Classical | No | | | | | No | No |
| V2214 Oph | Classical | No | 8.5 | 56 | 92 | 2.8204 | No | No |
| V2264 Oph | Classical | No | 9.9 | 19 | 45 | | No | No |
| N Pup 2004 (V574 Pup) | Classical | No | | 13 | < 27 | | No | No |
| V351 Pup | Classical | No | 6.4 | 10 | 26 | | No | No |
| V597 Pup | Classical | No | | | | | No | No |
| V598 Pup | Classical | No | | | | | No | No |
| T Pyx | Recurrent | No | 6.5 | | 88 | 1.829 | No | No |
| N Sco 2004b (V1187 Sco) | Classical | No | | | | | No | No |
| N Sco 2005 | Classical | No | | | | | No | No |
| N Sco 2007 (V1280 Sco) | Classical | No | | | 34 | | Yes | Yes |
| N Sco 2008 (V1309 Sco) | Classical | No | | | | | No | No |
| N Sco 2011b (V1313 Sco) | Classical | No | | 5.8-8.5 | 13-18 | | Yes | No |
| U Sco | Recurrent | Yes | 8 | | 7 | 29.53 | Yes | Yes |
| V745 Sco | Recurrent | No | 18.6 | | 14.9 | 12240 | No | No |
| V977 Sco | Classical | No | 9.4 | 3 | 8 | | No | No |
| V443 Sct | Classical | No | 8.5 | 19 | 28 | | No | No |
| V444 Sct | Classical | No | 10.5 | 6 | 10 | | Yes | No |
| N Sgr 2002c (V4743 Sgr) | Classical | No | 5 | | | 6.7 | No | No |
| N Sgr 2006 (V5117 Sgr) | Classical | No | | | | | No | No |
| V3890 Sgr | Recurrent | No | 8.4 | | 17 | 29.52 | No | No |
| V4157 Sgr | Classical | No | 7 | 10 | 22 | | No | No |
| V4160 Sgr | Classical | No | 7 | 6 | 14 | | No | No |
| N TrA 2008 (NR TrA) | Classical | Yes | 8.5 | | | 5.25 | No | No |
| CSS081007:030559+0547 | Classical | No | | | | | No | No |
| XMMU J115113.3-623730 | Classical | No | | | | 8.6 | No | No |

**References:** Ritter Cataclysmic Binaries Catalog, ed. 7.19 (Ritter & Kolb 2003); Stony Brook/SMARTS Atlas of (mostly) Southern Novae (Walter et al. 2012); CTIO Nova Survey (Williams et al. 1991, 1994).

**Table 2**

Input parameters for the RAINY3D simulations

| Nova | $q = M_2/M_1$ | $T_{eff}$ ($10^3$ K) | $L_*$ (erg s$^{-1}$) | $\log(g)$ (cm s$^{-2}$) | $\log(r_{in})$ (cm) | $\log(r_{out})$ (cm) | E(B-V) | $d$ (kpc) | *Ref.* |
|---|---|---|---|---|---|---|---|---|---|
| U Sco (cg) | 0.5 - 1.0 | 84 | 36.3 | 7.5 | 10.58 | 11.28 | 0.15 | 7.5 | 1 |
| U Sco (og) | 0.5 - 1.0 | 84 | 36.3 | 7.5 | 11.50-11.68 | 12.02-12.50 | 0.15 | 7.5 | 1 |
| V 394 CrA | 0.5 - 1.0 | 100 - 150 | 36.0 - 38.0 | 7.5 | 10.44 | 11.14 | 1.10 | 4.2 | 2 |
| KT Eri | 0.5 - 1.0 | 100 - 150 | 36.0 - 38.0 | 7.5 | 11.57 | 12.27 | 0.08 | 7.0 | 3 |

**Notes.** The U Sco (cg) parameters correspond to the values used in the closed geometry simulations and the U Sco (og) in the open geometry ones.

**References.** (1) Hachisu et al. 2000 ; (2) Hachisu & Kato 2000; (3) Imamura & Tanabe 2012.